\documentclass[aps,prl,twocolumn,floatfix,10pt]{revtex4-1}

\usepackage{amssymb}
\usepackage{amsmath}
\usepackage{graphicx}
\usepackage{subfigure}

\begin{document}

\title{Electronic Correlation and Transport Properties of Nuclear Fuel Materials}
\author{Quan Yin}
\email{Email: leoquan@gmail.com}
\author{Andrey Kutepov}
\author{Kristjan Haule}
\author{Gabriel Kotliar}
\affiliation{Department of Physics and Astronomy, Rutgers University, Piscataway, NJ 08854}
\author{Sergey Y. Savrasov and Warren E. Pickett}
\affiliation{Department of Physics, University of California, Davis, CA 95616}
\date{\today}

\begin{abstract}
The electronic structures and transport properties of a series of actinide mono carbides, mono nitrides and dioxides are studied systematically using a combination of density functional theory and dynamical mean field theory. The studied materials present different electronic correlation strength and degree of localization of $5f$--electrons, where a metal--insulator boundary naturally lies within. We also investigate the interplay between electron--electron and electron--phonon interactions, both of which are responsible for the transport in the metallic compounds. Our results allow us to gain insights into the roles played by different scattering mechanisms, and suggest how to improve their thermal conductivities.
\end{abstract}

\maketitle

When engineering fuel materials for nuclear power, important thermophysical properties to be considered are melting point and thermal conductivity. Uranium and plutonium oxide fuels used in very high temperature fast breeder reactors have very high melting points, but they suffer from poor thermal conductivity, because in these insulating oxides only lattice vibrations conduct heat. Hence attention is turning to metallic fuels for the new generation of reactors, such as uranium carbide and nitride \cite{nuclear_power}. Understanding the physics underlying transport phenomena due to electrons and lattice vibrations in actinide systems is a crucial step toward the design of better fuels. In this Letter we conduct a systematic theoretical study on the electronic structures and lattice dynamics of actinide compounds, and show that systems close to the Mott transition from the metallic side are the best option.

Thermophysical properties of solids are determined from their electronic structures, but in actinides they are not well described by the traditional approaches based on density functional theory (DFT) within its local density approximation (LDA) due to strong electronic correlation. In this study we use an advanced electronic structure method based on the combination of LDA and dynamical mean field theory (LDA+DMFT) \cite{DMFT_RMP2006}, which has proven success in describing such strongly correlated problems \cite{Pu-vol,Pu-phn,Pu-valence,MOX}.Our full--potential charge self--consistent implementation of LDA+DMFT described in Ref.\cite{LDA+DMFT} is based on the DFT program of Ref.\cite{wien2k}. For the impurity solver we use the continuous time quantum Monte Carlo (CTQMC) algorithm \cite{CTQMC1,CTQMC2}. For late actinides such as Pu and beyond, we use the less expensive vertex corrected one--crossing approximation (OCA) \cite{DMFT_RMP2006}, which is very accurate in these more localized systems. All calculations were performed in the paramagnetic phase.

\begin{figure}[t]
\centering
\includegraphics[width=0.9\columnwidth]{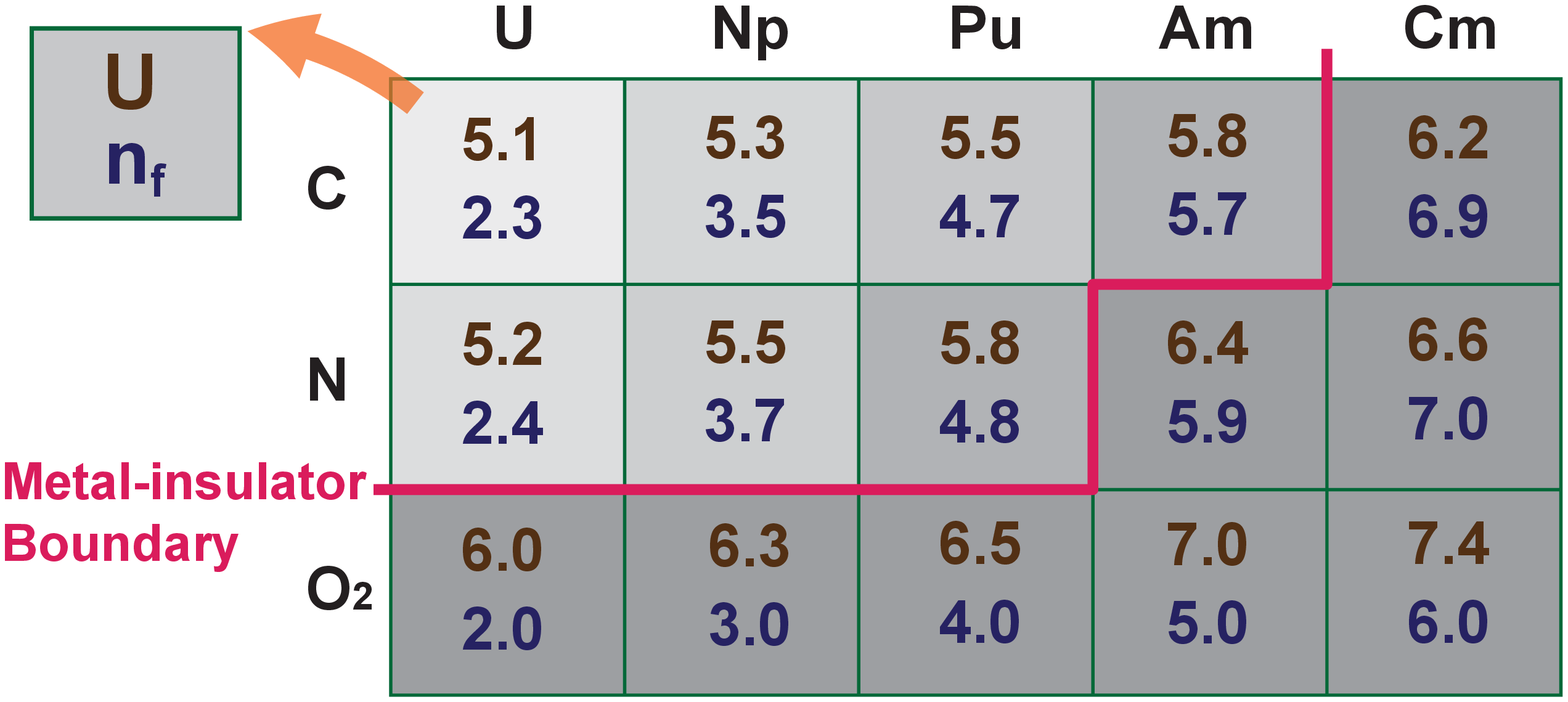}
\caption{Correlation diagram. The shading represents the
electronic correlation strength. The labels on the top denote the actinides elements, and the labels to the left denote the ligand elements. The red line is the metal--insulator boundary. Two quantities are listed in each cell: Hubbard $U$ (units: eV) and $f$-electron valence $n_{f}$.}
\label{fig:cor-diag}
\end{figure}

We first describe the chemical trends governing the degree of localization of the $f$--electrons in the binary actinide compounds listed in Fig. \ref{fig:cor-diag}. The key parameters are: the onsite Coulomb repulsion among the $5f$--electrons, quantified by the Hubbard $U$ and Hund's rule exchange $J$; the charge transfer energy $\Delta=\varepsilon_{f}-\varepsilon_{p}$; and the $5f$ band width quantified by the hybridization between $5f$ and $spd$ electrons. The charge transfer energy increases vertically from carbides to oxides due to the change in the electro--negativity of ligand atoms. The band width of $5f$-electrons shrinks horizontally from U to Cm compounds, indicating a more localized nature in late actinides. This causes a reduction of screening which is manifest in the gradual increase of $U$ from the left to the right, and from the top to the bottom of the table.

While most electronic structure methods can accurately calculate the hopping integrals between various electronic orbitals, evaluating the screened $U$ in solids is generally a difficult task. Here we have computed $U$ using a newly developed fully self--consistent many--body GW approach \cite{GW-Kutepov}, which provides a seamless interface with LDA+DMFT. The latter method allows to determine the degree of localization of the $5f$--electrons in each material. Our estimates for the Hund's $J$ are within the range of $0.5-0.6$ eV, about $30\%$ smaller than their atomic values due electronic screening. As a combination of the above quantities, the overall correlation strength and localization is visualized by the shading of Fig. \ref{fig:cor-diag}, referred as the ``correlation diagram" of binary nuclear fuel materials, where the gray gradient approximately represents the partial $f$ density of states at the Fermi level computed by LDA+DMFT.

\begin{figure}[t]
\centering
\includegraphics[width=1.0\columnwidth]{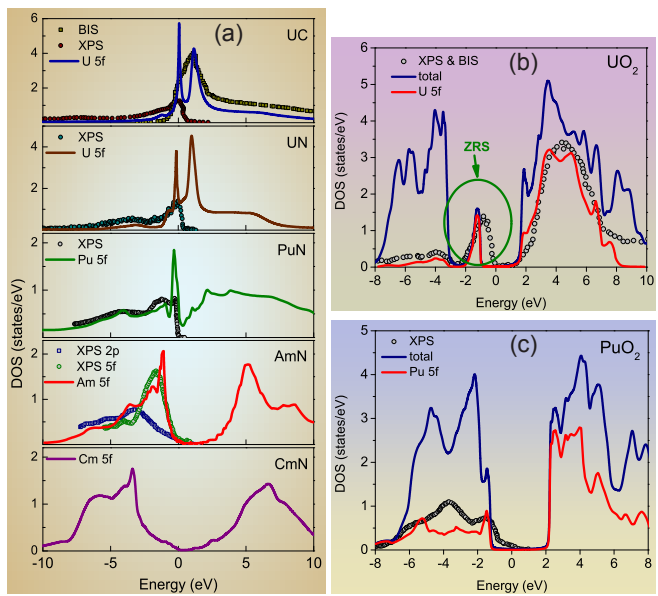}
\caption{Theoretical DOS compared with available x-ray photoemission spectroscopy. (a) Partial $5f$ DOS of UC and select actinide nitrides. The XPS \& BIS data of UC is from Ref.\protect\cite{XPS_UC}, UN from Ref.\protect\cite{UN_transport}, PuN from Ref.\protect\cite{XPS_PuN}, and AmN from Ref.\protect\cite{XPS_AmN}. (b)Total and partial DOS of UO$_{2}$. XPS and BIS taken from Ref.\protect\cite{XPS_BIS_UO2}. (c) Total and partial DOS of PuO$_{2}$. XPS from Ref.\protect\cite{XPS_PuO2}.}
\label{fig:dos}
\end{figure}

Next, we present the frequency dependence of the electronic spectral functions of some representative compounds in Fig. \ref{fig:dos}(a). From the top panel to the bottom, the $5f$ partial DOS changes qualitatively. UC and UN represent an itinerant $5f$--electron system with most spectral weight on the Fermi level, but the picture starts to change at PuN, where the Kondo resonance and satellite $5f$ states are present. In AmN the $5f$ DOS begins to form an marginal energy gap. The evolution of the density of states from UN to CmN echoes the itinerancy--localization transition of $5f$--electrons, and demonstrates the metal--insulator transition in a transparent point of view. CmC, CmN, and all the actinide oxides are also found to be insulators. This allows to establish a metal--insulator transition boundary, illustrated by the red line in Fig. \ref{fig:cor-diag}.

The actinides ions in most of the metallic crystals are found to be in a mixed valence state, where they do not settle in one valence, but fluctuate between different valences in the solid. It can be described by an effective number $n_{f}$ (listed in Fig. \ref{fig:cor-diag}), obtained using a valence histogram technique \cite{Pu-valence}, which represents an average over all the atomic configurations weighted by corresponding probabilities.

The total and partial DOS of UO$_{2}$ and PuO$_{2}$ are shown in Fig. \ref{fig:dos}(b) and \ref{fig:dos}(c). Most noticeably, the situation $U>\Delta$ allows us to describe the insulating actinide oxides as charge transfer Mott--insulators \cite{ZSA}, which is well known from late transition metal oxides, for example NiO, the classical textbook example of strongly correlated systems. Thus they could exhibit the Zhang--Rice state (ZRS) \cite{ZRS}, which is the low--energy resonance corresponding to the coupling of local moments of correlated electron orbitals to the hole induced by phototemission process on ligand orbitals. The appearance of the ZRS in UO$_{2}$ can be understood from the existence of a local magnetic moment in the U $5f^{2}$ many--body ground state, which is the $\Gamma _{5}$ triplet. It is also clear that PuO$_{2}$ does not show the ZRS because its ground state of the $5f^{4}$--shell is the $\Gamma _{1}$ singlet.

Experimentally, UC is a Fermi liquid (FL) at room temperature and ARPES measurement indicates that the overall band width is reduced by a factor of $4$ relative to the LDA band structure \cite{ARPES_UC}. In our calculation UC is a FL below $300K$ with $m^{\ast }/m_{LDA}=3.7$. On the other hand, UN shows a strongly correlated heavy fermion character with a coherence temperature below its Neel temperature of $53K$. In the absence of magnetic order, UN would be a FL at very low temperature with a large mass enhancement ($m^{\ast }/m_{LDA}\approx 12$) as can be inferred from the linear specific heat coefficient \cite{UN_spheat}. It is a non FL in the temperature range ($55-1000K$) we studied.

After understanding the electronic structures, we turn to the transport properties. We focus on correlated metallic compounds, where electrons play the role of charge and heat transporters, while retaining a high melting point. 
Although in normal metals electron--phonon scattering is dominant except at very low temperatures, in strongly correlated metals electron--electron scattering takes the lead. The electronic contribution to thermal conductivity is proportional to the electrical conductivity via the Wiedemann--Franz law. From the electronic structure and correlation strength of the studied materials, small resistivity occurs in the least correlated compounds in our table. Indeed UC and UN are the best fuel materials in terms of their outstanding transport properties.

Strong Coulomb interactions among electrons can substantially reduce the interaction between electrons and lattice vibrations \cite{Gun}. Hence the electron--phonon interaction (EPI) is usually weaker in strongly correlated materials, which might lead to smaller resistivity due to EPI. On the other hand, increasing electronic correlations leads to an increase in resistivity due to electron--electron scattering. Therefore neither extremely weak nor strong correlations is good from the perspective of minimizing resistivity. Deciding 
the optimal degree of correlation for the purpose of maximizing conductivity thus requires first--principle calculations. 

\begin{figure}[t]
\centering
\includegraphics[width=0.75\columnwidth,draft=false]{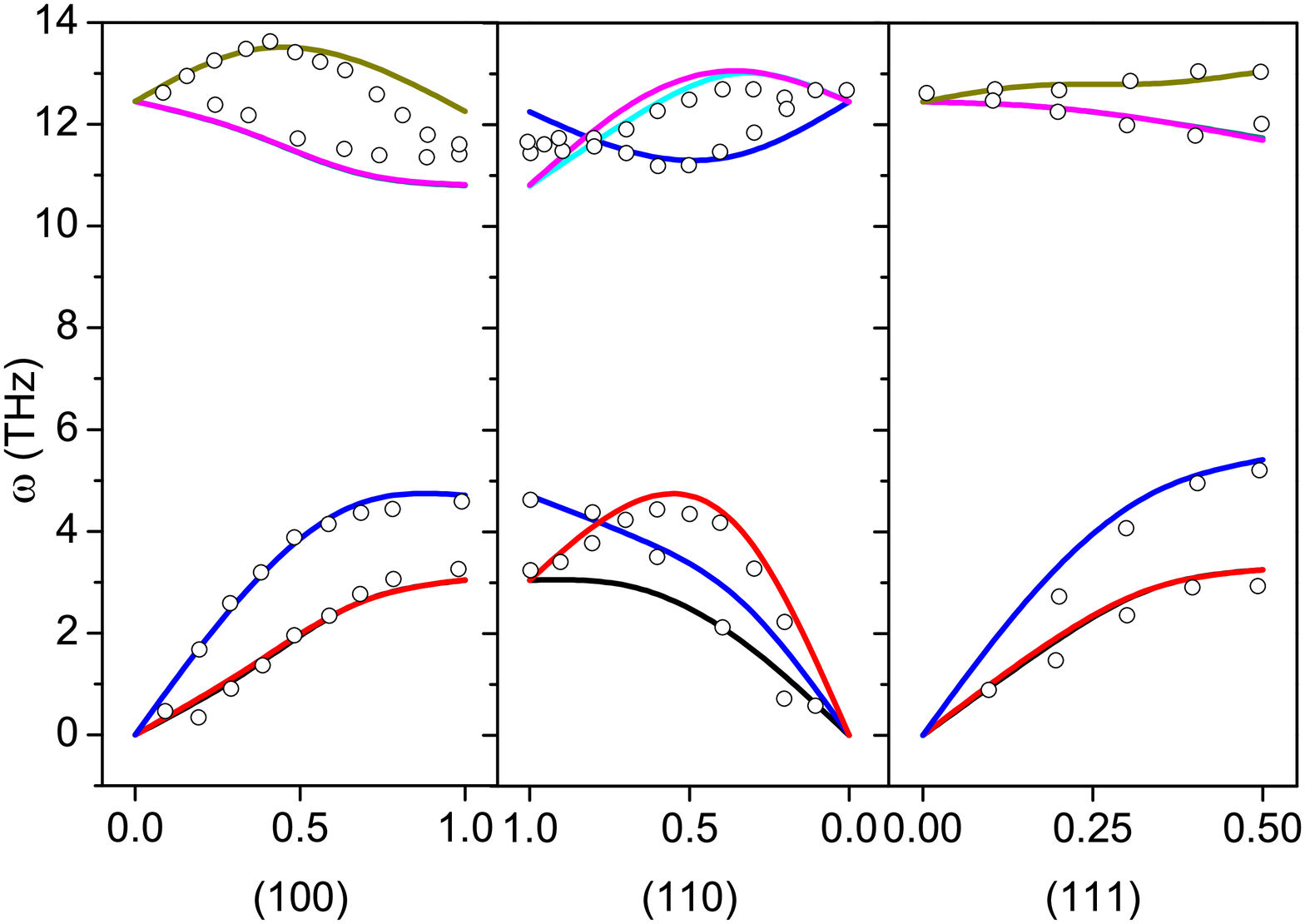}
\caption{Phonon dispersions of UN plotted along three high--symmetry lines. Solid curves: Calculated phonon dispersion. Open circles: Experimental
phonon excitations from Ref.\protect\cite{UN_phonon}.}
\label{fig:phn}
\end{figure}

For the calculation of phonon spectra we used the density functional linear
response approach implemented in the LMTO basis \cite{lmtart,LR-LMTO}. The spin--orbit coupling effect is included fully relativistically. The phonon dispersion of UN along $3$ high--symmetry directions is plotted in Fig. \ref{fig:phn} together with experimental data measured by neutron scattering \cite{UN_phonon}. UC carries very similar phonon dispersions but slightly lower phonon energies. Despite apparent presence of correlation effects, excellent agreement is achieved with the local density approximation (LDA). Similar success of LDA in studying lattice dynamics of strongly correlated metallic systems have been reported earlier, for example in cuprates \cite{HTSC}, and recently iron pnictides \cite{LILIA}.

Calculations of electron--phonon interactions and transport properties
require, on the other hand, quasiparticle description of the one--electron
spectra when evaluating Eliashberg and transport spectral functions by integrating over the Fermi surfaces \cite{e-p_transport}. As a result, due to
large mass enhancement, the straightforward LDA procedure can produce wrong electron--phonon resistivities, which was indeed found in our calculation for UC where $\rho(T)_{EPI}$ was overestimated by a factor of $3$ compared to experiment.

In order to evaluate the electron--phonon scattering in the presence of
correlations, we develop a method that accounts for the effects from quasiparticle mass renormalizations and spectral weight transfer by utilizing interacting Green's functions. The use of the pole interpolation of self--energy allows us to replace the non--linear (over energy) Dyson equation by a linear Schroedinger--like equation in extended subset of ``pole states" \cite{AmPRL}.
The advantage of this method is that the well developed machinery of standard electronic structure methods can be simply generalized to account for the dynamical self-energy effects.

In the present work we extend this method to compute the electron--phonon
interactions for UC and UN, whose $f$--electrons show itinerant behavior with $m^{\ast }/m_{LDA}\approx 4-12$. To capture this mass renormalization effect, we first make a fit to the self--energy obtained from the CTQMC, using a two--pole interpolation where the slope of the self--energy at zero frequency $d\Sigma (\omega )/d\omega |_{\omega=0}=1-m^{\ast }/m_{LDA}$ controls the electronic mass enhancement, and the positions of the two poles determine the transfer of the spectral weight from the quasiparticle band to the Hubard bands. Next, we assume that the $f$--electrons are rigidly bound to their ions so that there is no actual change in the self--energy, caused by ion displacements from their equilibrium positions. Since the main contribution to electronic transport comes from the states near the Fermi surface, where quasiparticles are best described in terms of slave bosons, the neglecting of $\delta \Sigma (\omega )$ due to ion displacements corresponds to a rigid self--energy approximation. This is very similar to the famous rigid muffin--tin approximation (RMTA) \cite{RMTA}, which has been successfully applied in the past to study electron--phonon interactions in transition metal compounds \cite{PAPA,Mazin}. The electron--phonon scattering matrix element $g_{\mathbf{k}j,\mathbf{k}+\mathbf{q}j^{\prime }}$ is then evaluated using the electronic components $\psi _{\mathbf{k}j}^{(e)}$ that appear as solutions to
the extended Schroedinger--like equation. These matrix elements can be subsequently used for evaluating the EPI part of electrical and thermal resistivity \cite{e-p_transport}, where the corresponding Fermi surface integrals are now performed with the extended ``band structures" $\epsilon _{\mathbf{k}j}^{(e)}$ that acquire renormalizations due to correlations. Using this method, a reduction (by a factor of $3$) in $\rho (T)_{EPI}$ for UC was obtained, while in UN the effect was marginal.

To evaluate the conductivity due to electron--electron scattering we use the Kubo formalism \cite{LDA+DMFT}, where the scattering rate comes from the imaginary part of DMFT self--energy $\Sigma(\omega ,T)$, obtained from CTQMC. 

\begin{figure}[t]
\centering
\subfigure[]{\includegraphics[width=0.75\columnwidth,draft=false]{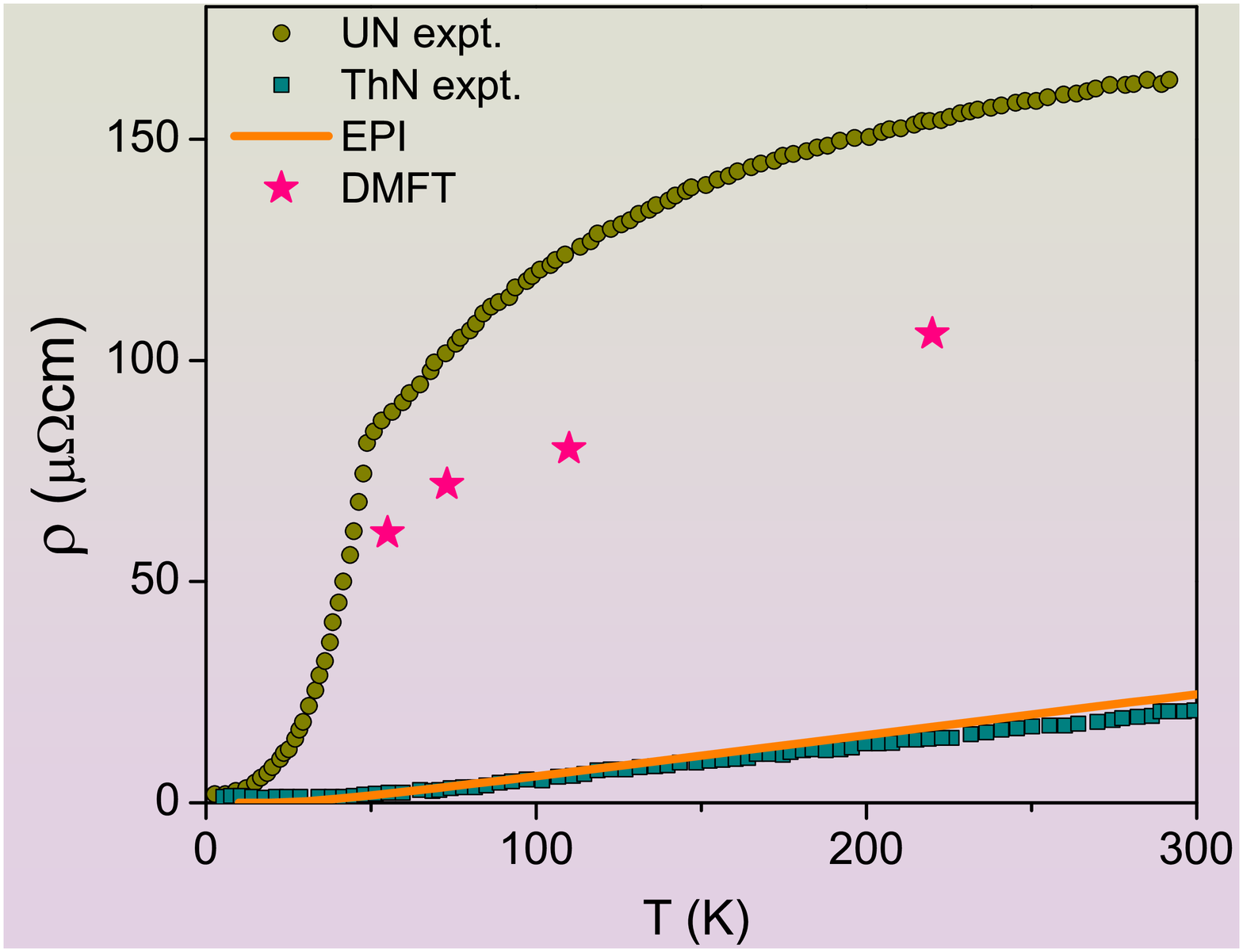}}
\subfigure[]{\includegraphics[width=0.75\columnwidth,draft=false]{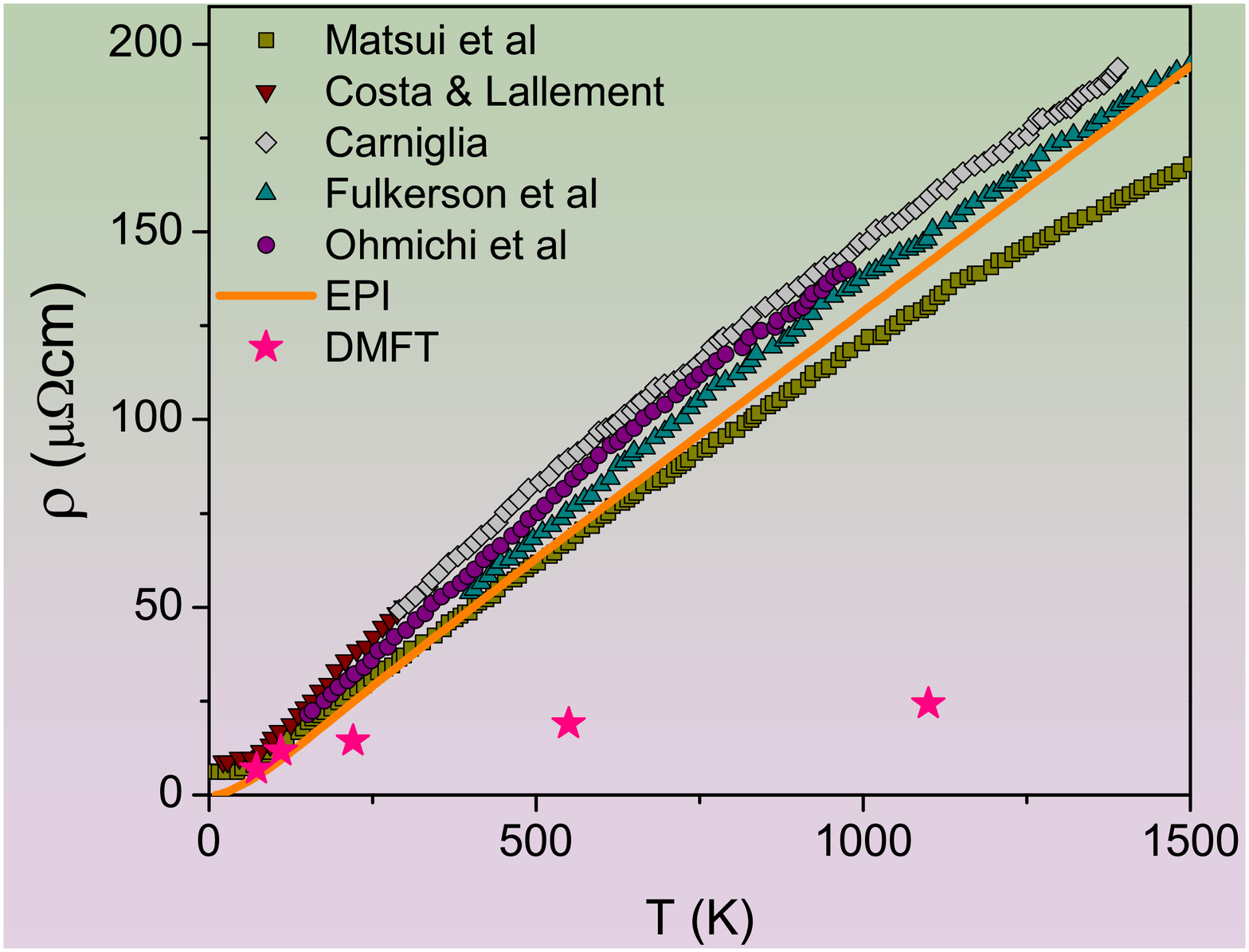}}
\caption{Electrical resistivity due to two different scattering mechanisms. The electron--phonon interaction (EPI) resistivity is show as solid lines, and electron--electron interaction resistivity, which is calculated by LDA+DMFT (CTQMC) at several temperatures, is shown as stars. (a) UN. Experimental resistivity data are taken from Ref.\protect\cite{UN_transport}. (b) UC. Experimental data after Ref.\protect\cite{UC_res_1,UC_res_2,UC_res_3,UC_res_4,UC_res_5}.}
\label{fig:res}
\end{figure}

We can now build the entire picture of the electronic transport in the uranium compounds with our results summarized in Fig. \ref{fig:res}. Electron--electron scattering can account for approximately $80\%$ of $\rho(T)$ in UN, commonly found in heavy fermion systems, entitling it as a strongly correlated bad metal. In contrast, UC shows nearly linear $\rho(T)$, which is an indication of dominant electron--phonon scattering, and our calculated results indeed show that in UC, $\rho(T)_{ee}$ is much smaller than $\rho(T)_{EPI}$. Our findings verify the distinct characters in the electrical transport of UC and UN, two seemingly similar materials. By making a side--to--side comparison, we see that UC is less correlated than UN, which makes $\rho(T)_{ee}$ larger in the latter. Nevertheless $\rho(T)_{EPI}$ is larger in the former.

We also estimate lattice vibrational contribution to thermal conductivity in UC and UN. This is done by evaluating the Gruneisen parameter and phonon group velocities using the method described previously for oxide fuels \cite{MOX}. At $T=1000K$, lattice thermal conductivity $\kappa_{ph}=2.7W/mK$ in UC, and $\kappa_{ph}=4.4W/mK$ in UN, much smaller than electronic counterpart. Thus $\kappa_{ph}$ only plays a minor role in these two metallic uranium compounds.

We put together our results and evaluate total thermal conductivity at $1000K$, a representative temperature under which nuclear reactors operate. By applying the Wiedemann--Franz law on the electrical conductivity data, we obtain $\kappa_{ee}$. Since electronic thermal resistivity consists of two scattering processes, total thermal conductivity is estimated by $\kappa_{total}=(\kappa_{ee}^{-1}+\kappa_{EPI}^{-1})^{-1}+\kappa_{ph}$, in which the first two terms correspond to $\kappa _{electron}$. For UN, our result is $\kappa _{total}=16.5W/mK$, compares well with a recent study which extracted the phonon contribution from  molecular dynamics (MD)\cite{UN_MD} and the electronic contribution from experiments. Experimentally, $\kappa (1000K)\approx 19-23W/mK$. In UC, we obtained $\kappa _{total}=18.7W/mK$, also close to the experimental value of $23W/mK$ \cite{UC_kappa_expt}. The discrepancy between theory and experiment is likely due to other excitations that can conduct heat and are not accounted for in our calculation, and the approximate nature of the Wiedemann Franz law and Boltzmann transport theory which are used to obtain the electronic and lattice thermal conductivity, respectively.

At last, the understanding gained from our computational study suggests avenues for improving the thermal conductivity of UC and UN. Since optimizing thermal conductivity is equivalent to minimizing resistivity at high temperatures, it is interesting to look at the doping dependence of the resistivity in UC/UN, or in the solid solution UC$_{1-x}$N$_{x}$. Let us represent the total resistivity of UC$_{1-x}$N$_{x}$ by $\rho (T,x)=$ $\rho (T,x)_{ee}+\rho (T,x)_{EPI}$. Note that on one hand, $\rho (T,x)_{ee}$ should show a rapid growth when $x$ approaches $1$ since  $\rho(T)_{ee}\propto 1/T_{K}\propto W/Z$ above coherence temperature, where $T_{K}$ is the Kondo temperature and $W$ is the band width. On the other hand, $\rho (T,x)_{EPI}$ being proportional to transport coupling constant $\lambda _{tr}$ divided by the average square of the Fermi velocity $N(0)\langle v^{2}\rangle $ of the quasiparticles decreases with $x$ in our calculation. 
Therefore there exists a region between UC and UN where the total resistivity is minimized. It is also possible to achieve similar effects in UC by electron doping, or in UN by hole doping.

To conclude, we have carried out the first LDA+DMFT exploration of the electronic structure and transport properties of nuclear fuels. The actinide dioxides are charge--transfer insulators, where the Zhang--Rice state is present in UO$_{2}$. The metallic carbide and nitride compounds exhibit strong electronic correlations, which is reflected in the incoherent non Fermi
liquid behavior at temperatures relevant for nuclear reactions. We have
achieved a successful theoretical description of the transport in UC and UN. While UN clearly shows a strongly correlated signature, both the electron--electron and electron--phonon scattering mechanisms contribute to transport in the less correlated sister compound UC. Our findings enable us to give predictions on how to improve these two uranium based nuclear fuel materials.

\textbf{Acknowledgements} This work is supported by the United States Department of Energy Nuclear Energy University Program, contract No. 00088708. The authors acknowledge the support of Computational Materials Science Network, Grant NO. DE-FG03-01ER45876, and Grant NO. DE-SC0005468 linking UC Davis and Rutgers. WEP acknowledges DOE Grant NO. DE-FG02-04ER46111. QY thanks Viktor Oudovenko at Rutgers University for technical support on computation.

\end{document}